\documentclass{article}
\begin{document}

\begin{center}{\bf\large 
Thermodynamics of Extended Gravity Black Holes}

\bigskip

Guido Cognola$^{a}$, Lorenzo Sebastiani$^{a}$, Sergio Zerbini$^{a}$

\medskip

$^a$Dipartimento di Fisica, Universit\`{a} degli Studi di Trento,\\
Trento, 38100, Italy\\
INFN - Gruppo Collegato di Trento\\

\end{center}


\begin{abstract}
Thermodynamics of extended gravity static  spherically symmetric black hole solutions is investigated. The energy issue is discussed making use of the derivation of Clausius relation from equations of motion, evaluating the black hole entropy by the Wald method and computing the related Hawking temperature.
   
\end{abstract}





\section{Beyond General Relativity} 
To begin with we recall that recent astrophysical data are in agreement with a 
universe in current phase of accelerated expansion, in contrast with the predictions of Einstein gravity in FRW  space-time.
Most part of energy contents, roughly $75\, \%$ in the universe is due to  mysterious entity  with negative 
pressure: Dark Energy. The simplest explanation is  Einstein gravity plus a small positive cosmological constant.
As an  alternative, one may consider more drastic modification of General Relativity: Extended  Gravity Models (see, for example 
\cite{Capo,turner,No,fara,ea,seba08}). Thus, it is of some interest to investigate the existence of static spherically symmetric black hole solutions and the associated thermodynamical properties, in particular the non energy issue associated with energy. 
\section{First example: the Lovelock gravity} 
 Lovelock gravity is an extended gravity, with higher derivative terms but in higher dimensions, and it deals only with second order partial differential equations in the metric tensor as in GR. Lovelock Lagrangian contains  Euler densities
\begin{equation}
  {\cal L}_m = \frac{1}{2^m} 
  \delta^{\lambda_1 \sigma_1 \cdots \lambda_m \sigma_m}_{\rho_1 \kappa_1 \cdots \rho_m \kappa_m}
  R_{\lambda_1 \sigma_1}{}^{\rho_1 \kappa_1} \cdots  R_{\lambda_m \sigma_m}{}^{\rho_m \kappa_m}
\end{equation}
where  $R_{\lambda \sigma}{}^{\rho \kappa}$  Riemann tensor in $D$-dimensions
and $\delta^{\lambda_1 \sigma_1 \cdots \lambda_m \sigma_m}_{\rho_1 \kappa_1 \cdots \rho_m \kappa_m}$  
generalized totally antisymmetric Kronecker delta. The action for Lovelock gravity is
\begin{equation}
I=\int d^Dx \sqrt{-g}\left[-2\Lambda+\sum_{m=1}^k\left\{\frac{a_m}{m}{\cal L}_m\right\}\right]\,,
\end{equation}  
 $k\equiv [(D-1)/2]$ and  $a_m$ are arbitrary constants. 
Here, $[z]$ represents the maximum integer satisfying $[z]\leq z$, and $a_0=-2\Lambda$ and $a_1=1$.  
For $D=5$, $n=3$, the spherically symmetric static solution reads \cite{ldeser} 
\begin{equation}
ds^2=-B(r)\,dt^2+\frac{dr^2}{B(r)}+r^2d^2S_3\,,
\end{equation}
with 
\begin{equation}
B(r)=1+\frac{1}{a_2}\pm \frac{1}{a_2}\sqrt{1+2a_2(\frac{C}{r^2}+\frac{\Lambda}{3}r^2)}
\end{equation}
$a_2$ the Gauss-Bonnet parameter. There exists a BH solution as soon as $B(r_H)=0$, for $r_H>0$. Following, for example \cite{seba}, making use of Killing vector  $K_\mu$, $ \nabla_\nu K_\mu+\nabla_\mu K_\nu=0$, one can construct the conserved current
$J_\mu={\cal G}_{\mu \nu}{}K^{\nu}\,, \quad
\nabla_\nu J^{\nu}=0\,,$
where ${\cal G}_{\mu\nu}={\cal G}_{\nu\mu}$ is the conserved Einstein-Lovelock tensor. The quasi-local generalized Misner-Sharp energy is the conserved charged of the current  $J_\mu$, namely
\begin{equation}
E(r)=-\frac{1}{8\pi}\int_\Sigma d \Sigma_\mu K_\nu{\cal G}^{\mu \nu}=\frac{3V(S_3)}{16\pi}r^{4} W(B(r))\,,
\end{equation}
where $\Sigma$ is a spatial finite volume at fixed time and with $0< r'<r$, $d\Sigma_\mu=(d\Sigma, \vec 0)$. $W(B)$ depends on $B(r)$. 
For example,  in $D=4$, Lovelock gravity is General Relativity plus $\Lambda$, and one has
$
E(r)=\frac{r}{2}\left((1-B(r))-\frac{\Lambda}{3}r^2\right)\,.
$
On shell, Misner Sharp BH mass is $E=\frac{3V(S_3)}{16\pi}C$, namely BH  energy is proportional to the constant of integration $C$.
On the other hand, well know methods of  QFT  give  for  the Hawking temperature $T_H=\frac{B'_H}{4\pi}$. Furthermore, one can compute BH 
entropy via well known Wald method . As a result
\begin{equation}
T_H\,dS_W=dE=\frac{3V(S_3)}{16\pi}dC \,,
\end{equation}
namely, for Lovelock gravity, the Clausius Relation holds, as in GR.
\section{The generic extended gravity models}

What about other extended  gravity models? In general, it is rather difficult to define  the analogue of the quasi-local Misner-Sharp mass. However, in $D=4$, for BH energy related to a class of extended gravity, one may make use of the following procedure  \cite{gorbunova}: assume to know a BH exact solution with $r_H$ depending only on a unique constant of integration $C$. Compute the Wald entropy: if the black hole Entropy depends 
only on $r_H$, then make use of the knowledge of $T_H$ and of the Clausius Relation $T_HdS=dE$. As a result,  BH energy turns out  proportional to $C$.  This proposal  is supported by the  derivation of the Clausius Relation from the equations of motion, and by evaluation of the BH entropy via Wald method and the Hawking temperature via the quantum mechanics techniques in curved space-time.  
\section{Second example: a class of four-dimensional $F(R)$  modified gravity models}
The action is 
\begin{equation}
I=\int d^4 x\sqrt{-g}F(R)\,, 
\end{equation}
 $F(R)$ is a generic function of the Ricci scalar $R$. The Metric Ansatz for  BH solutions 
\begin{equation}
ds^2=-B(r)a(r)^2\,dt^2+\frac{dr^2}{B(r)}+r^2dS^2_2\,.
\end{equation}                                                                                                                                     In general $a(r)$ is non trivial. Again BH solutions exist when   $B(r_H)=0$ with $r_H>0$.  As already mentioned, the two relevant quantities are:  the  Hawking temperature 
\begin{equation}
T_{H}=\frac{\kappa_H}{2\pi}=\frac{a(r_H)}{4\pi} B'_H \,,
\end{equation} 
 and  the BH entropy evaluated by   by the Wald method, which reads
\begin{equation}  
S_W=\frac{\mathcal{A}_H}{4}f(R_H)\,,\quad f(R)=\frac{d F(R)}{d R}\,. 
\end{equation} 
Note that, once the BH solution is known,  $T_H$ and $S_W$ can be evaluated.
Now assume that $r_H$ depends only on a constant of integration $C$ and $R_H=R_H(r_H)$.  From  equations of motion on horizon and Wald entropy and $T_H$, one gets
\begin{equation}
T_{H}dS_W = a(r_H)\left(\frac{f_H}{2}-\frac{R_Hf_H-F_H}{4 }r_H^2\right)dr_H\,.
\end{equation} 
Interpretation: this is the Clausius Relation. Thus, integrating over $r_H$, one gets the  $F(R)\,\, BH $ energy formula n\cite{seba}     
\begin{equation}
 E=\int\, a(r_H)\left(\frac{f_H}{2}-\frac{R_Hf_H-F_H}{4 }r_H^2 \right)dr_H \,.
\end{equation}
 Several  exact BH  solutions are known, and one has a non trivial check of above $F(R)$ BH  energy expression  and $E \equiv C$.
For example, the Clifton-Barrow solution associated with $F(R)=R^{1+\delta}$ (see, for example \cite{clifton,Bel}), has been shown to satisfy  the 
above energy formula.




\end{document}